\documentclass[twocolumn,prb,color,psfig,showkeys,superscriptaddress]{revtex4}

\usepackage{graphicx}
\usepackage{amsmath}
\usepackage{amssymb}

\newcommand{\eps}{\varepsilon}
\newcommand{\s}{\sigma}
\newcommand{\w}{\omega}
\newcommand{\lam}{\Lambda}

\begin{document}

\title{Optical conductivity of Bose droplets in quenched disorder at $T = 0$:
the effects of structure factor.}

\author{E.V. Zenkov}
\email{eugene.zenkov@mail.ru}
\affiliation{Ural State Technical University, 620002 Ekaterinburg, Russia}

\begin{abstract}
The optical response of a system of two-dimensional Bose condensate droplets with an
internal charge structure interacting with impurities is considered within the
mode-coupling theory. It is found, that the glass transition in the system of droplets
occurs at much stronger disorder as compared to the case of stuctureless point
particles. The optical conductivity of the system is found to exhibit a complex
oscillating structure, intimately related to the charge distribution within the
droplets. The application of the model to the problem of the far IR optical
conductivity of doped cuprates La$_{2-x}$Sr$_x$CuO$_4$ ($x < 0.26$) is discussed.
\end{abstract}


\maketitle

\section{Introduction}

In a recent paper \cite{epl}, a simple model of the superfluid-glass transition in the
system of 2D bosons at zero temperature have been presented. The exact analytic
solution of the self-consistent equation for the memory function permitted to retrace
the quantum phase transition of the Bose condensate from superfluid to the insulating
(Bose-glass) regime, as the dimensionless parameter, measuring the strength of the
short-range random potential of impurities (pinning centers) reaches some critical
value.

In the present article we extend this model to consider the behaviour of the
disordered system of Bose droplets  with some internal charge structure, interacting
with random impurities. The droplet may extend over a few unit cells on a crystalline
layer and is a region of a quasi-free motion of a Bose particle. It may be regarded as
a potential well, where the boson is confined, and the radius of the droplet
corresponds to the relevant localization length. The droplet acts as a large heavy
Bose quasiparticle (like a large polaron). To our mind, a promising physical example
of the droplets are the local islands of superconducting phase of HTSC materials above
$T_c$ (see e.g. Ref.\cite{iguchi}), which existence is sometimes associated with
charged stripes \cite{Markiewicz}. Our approach is based on the mode-coupling theory
of diffusion and localization \cite{gotze}, that makes it possible to derive the
memory function of the system, given the bare correlation functions and the parameters
of the random potential.

The main qualitative result of our analysis is that the internal real-space structure
of the droplet strongly affects the electromagnetic response of the system and gives
rise to the novel features in optical conductivity spectra as compared to the case of
structureless point Bose particles. In the next section we formulate the model and in
Section 3 show the results. In Section 4 the applications of the model to the
experimental data on microwave optical response of doped cuprates are discussed. We
believe, that the collective density modes in the system of the Bose droplets may give
rise to the features in the optical conductivity spectra, that are usually ascribed to
the dynamics of the charged stripes.

\section{The model}

We consider the two-dimensional (2D) system of heavy Bose particles at $T = 0$,
randomly distributed on a background with real static dielectric constant
$\varepsilon$. Each particle is a droplet, carrying a charge, spread over the droplet
with the density $\rho(\mathbf{r})$. The interaction between the droplets is taken
into account within the random phase approximation, assuming, that the system is
dilute enough and the droplets do not overlap.

Our goal is to study the optical response of the system in presence of the random
impurity potential. The main quantity of interest in this problem is the memory
function $M(k, z) = M^{\prime} + i\,M^{\prime\prime}$, that makes it possible to write
down the optical conductivity of the system in a general form:
\begin{equation}\label{drude}
\s^{\prime}(z) = \frac{\w_p^2\,M^{\prime\prime}}{(z + M^{\prime})^2\,+\,M^{\prime\prime 2}}
\end{equation}
where $\w_p$ is the plasma frequency, $z$ -- the complex (Laplace) frequency. As
usual, the $k$-dependence is omitted in eq.(\ref{drude}), so that the memory function
evaluated at $k = 0$ is of interest. Depending on the analytic properties of $M(z)$,
eq.(\ref{drude}) yields the coherent optical response of a clean system, or describes
the diffusive charge dynamics in a dirty localized phase. The common practical method
of exploring the extended-localized phase transition in disordered system is provided
by the mode-coupling theory \cite{gotze,goldBose}, that yields the self-consistent
equation for the memory function:
\begin{equation}\label{m1}
M(z) = i\,\gamma + \frac{1}{d\,n\,m} \int\limits_{0}^{\infty} \Phi\left(z + M(z);
k\right) \langle |U(k)|^2 \rangle k^2 d\mathbf{k},
\end{equation}
where $\gamma$ is a bare relaxation constant, $\Phi$ is the density-density relaxation
kernel \cite{goldBose}, $U(k)$ is the Fourier component of the impurity potential,
$d$, $n$, $m$ are the dimensionality of the system, the concentration and mass of the
constituent charge carriers, respectively. The angular brackets $\langle\hdots\rangle$
stand for the average over the configurations of impurities. In the case of random
configurations this reduces merely to the factor $n_i$, the concentration of
impurities. It may be seen, that in absence of impurities $M = i\,\gamma$ and the
optical conductivity, eq.(\ref{drude}), reduces to the simple Drude form. The
approach, based on eqs.(\ref{drude},\ref{m1}) have been applied to a number of model
systems, both classical \cite{yip} and quantum Fermi \cite{gotze} and Bose
\cite{goldBose} fluids.

Denoting the random impurity potential as $U(r) = U_0 \varphi(r)$, where $\varphi(r)$
is dimensionless, the properly scaled equation (\ref{m1}) for the memory function
$M(0, z)$ of 2D Bose liquid at $T = 0$ K takes the form \cite{note}:
\begin{equation}\label{m2}
M\,=\,\frac{8 \Lambda}{\pi}
 \int\limits_{0}^{\infty}
\frac{\left(z + M\right) \varphi(k)^2\, k^{\nu + 3}}{(1 + k^{2 + \nu})\,\left(k^4 + k^{2 - \nu} - z (z + M)\right)}\,dk,
\end{equation}
where $\nu = 1$ for Coulomb interparticle interaction, $V(r) \sim 1/r$, and $\nu = 2$
for logarithmic (''2D Coulomb'') interaction, $V(r) \sim \log(r)$. This latter type of
interaction is typical of the topological extended structures, such as the vortices in
the type II superconductors. The ground-state properties of the 2D Bose gas with
logarithmic interaction have been considered in Ref.\cite{tanatar}. All quantities in
eq.(\ref{m2}) are dimensionless, the units of momentum and energy being:
\begin{equation}\label{units}
p_0 = \left(\frac{8\,\pi\,\hbar\,m^*\,n_{2D}\,Q^2}{\eps}\right)^{\frac{1}{2 + \nu}},\quad
\varepsilon_0 = \frac{p_0^2}{2\,m^*},
\end{equation}
respectively. Here $m^*$ is the effective mass of the droplet, $Q$ -- the charge,
$n_{2D}$ -- their concentration and $\eps$ is the real static dielectric constant of
the medium. The properties of the memory function are governed by a dimensionless
parameter
\begin{equation}\label{lamda}
 \Lambda = \frac{\pi}{2} \frac{n_i}{n_{2D}} \left(\frac{U_0}{\eps_0}\right)^2 =
         \frac{n_i\,m}{4 n_{2D}^2}\left(\frac{U_0}{\hbar  Q}\right)^2,
\end{equation}
where $n_i$ is the concentration of impurities.

Taking into account the internal structure of the droplet, the impurity potential
$\varphi(r)$ should be averaged with its charge density $\rho(r)$. Hence, the Fourier
transform $\varphi(k)$ is multiplied by the form-factor $\rho(k)$ of the droplet.
Thus, new expression
\begin{equation}\label{S}
 \widetilde{\varphi}(k)^2 = \varphi(k)^2 S(k)
\end{equation}
is to be put into eq.(\ref{m2}) instead of $\varphi(k)^2$, where $S(k) = |\rho_k|^2$
is hereafter called the structure factor. The form-factor of an isotropic 2D droplet
is calculated using the formula:
\begin{equation}\label{bessel}
 \rho_k  = 2 \pi \int \limits_0^{\infty} \rho(r) J_0(k r) \,r\, dr,
\end{equation}
where $J$ is the Bessel function of the first kind.

\section{Results}

\subsection{Superfluid-glass transition}

The model under consideration allows full analytic treatment \cite{epl} for the system
of point 2D logarithmic droplets ($S(k) = 1$, $\nu = 2$) interacting with pinning
centers -- the random impurities with the short-range potential $\varphi(k) = 1$. The
electromagnetic properties of such a model share much similarity with those of pinned
charge density waves (CDW) \cite{fukuyama}. The basic qualitative results, that apply
to a broad range of superfluid-glass transition models \cite{goldBose,goldHTS}, are as
follow.

\begin{figure*}[t]
\begin{minipage}[b]{0.48\linewidth}
\includegraphics[width=\linewidth,angle=0]{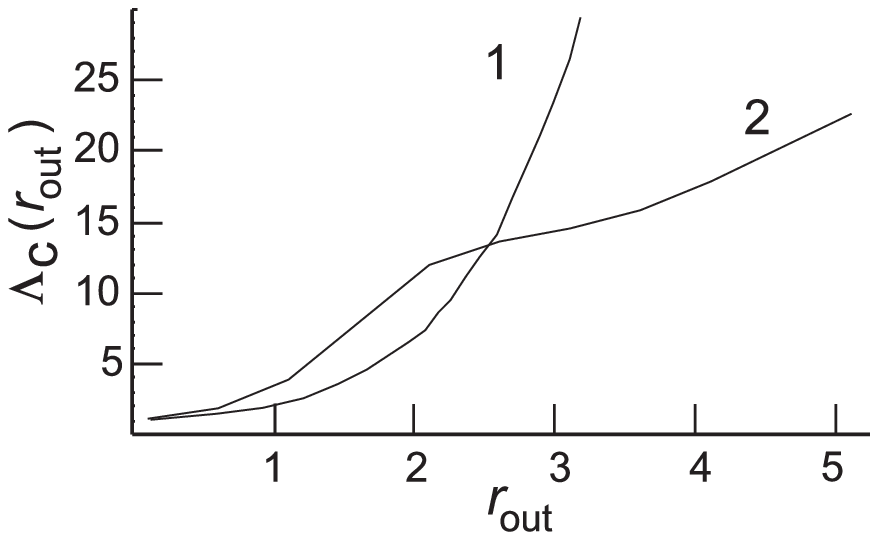}
\end{minipage}\hfill
\begin{minipage}[b]{0.48\linewidth}
\includegraphics[width=\linewidth,angle=0]{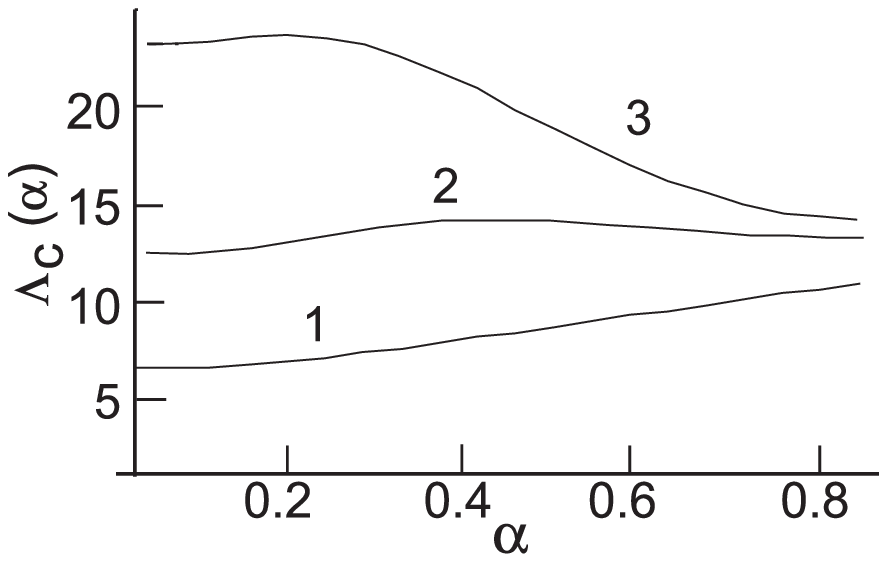}
\end{minipage}\hfill
\caption{Left panel: The dependence of the critical disorder parameter $\lam^*$,
eq.(\ref{onset}), on the outer radius of the droplet for some inner-to-outer radius
ratio $\alpha$: 1 -- $\alpha = 0.01$, 2 -- $\alpha = 0.99$.  Right panel: The dependence
$\lam^* (\alpha)$ for different $r_{out}$: 1 -- $r_{out} = 2.0$, 2 -- $r_{out} = 2.5$, 3
-- $r_{out} = 3.0$.} \label{Lc}
\end{figure*}

Until $\lam$ is small, the low-frequency expansion of $M(z)$ is:
\begin{equation}
 M \simeq m_1 z + \hdots,
\end{equation}
($m_1 = const.$), and the optical conductivity (\ref{drude}) is singular: $\s^{\prime}
= \frac{\w_p^2}{1 + m_1}\,\delta(\w)$, where $z = \w + 0 i$. Thus, the system is
superfluid at small disorder. However, as the full spectral weight $\int_0^{\infty}
\s^{\prime}(\w) d\w$ is constant, the reduction of the $\delta(\w)$-response by a
factor $1 + m_1$ as compared to the case of free Bose-condensate entails the emergence
of the spectral features at $\w > 0$. The spectral weight of the $\delta$-peak at zero
frequency drops with increasing $\lam$ and it vanishes at a critical value $\lam =
\lam^*$. At this point, the memory function acquires the singularity, called the
non-ergodicity pole, and in leading order in $z$:
\begin{equation}\label{nonerg}
 M \simeq \frac{m_0}{z} + \hdots.
\end{equation}
In this regime, the optical conductivity tends to zero in the limit of small $z$, i.e.
the system is in insulating (glassy) phase. The onset of the localization corresponds
to the condition: $m_0 \rightarrow 0$. Substituting  this into
eqs.(\ref{m2},\ref{S},\ref{nonerg}) with $z$ tending to zero, one obtains the
following equation for $\lam^*$:
\begin{equation}\label{onset}
  \frac{8 \Lambda^*}{\pi} \int \limits_{0}^{\infty}
  \frac{\varphi(k)^2 S(k)}{(1 + k^{2 + \nu})^2} k^{2 \nu + 1} dk = 1.
\end{equation}

The solutions of eq.(\ref{onset}) for point particles ($S = 1$) are summarized in
Table \ref{lc}. In particular, for the model of 2D bosons with logarithmic interaction
($\nu = 2$) in short-range impurity potential ($\varphi = \delta(r)$) we recover the
condition $\lam^* = 1$, obtained in Ref.\cite{epl}. Since $\lam^*$ varies depending on
the model parameters, it is natural to make use of a renormalized invariant constant
$\lam/\lam^*$.
\begin{table}
\caption{$\lam^*$ for point Bose particles ($S(k) = 1$) derived from eq.(\ref{onset})
for different impurity potentials $\varphi$ and interparticle interactions ($\nu = 1$
-- Coulomb, $\nu = 2$ -- logarithmic).} \label{lc}
\begin{center}
\begin{tabular}{cccccc}
\hline
   $\varphi(r)$ & \multicolumn{2}{c}{$\delta(r)$} & & \multicolumn{2}{c}{$1/r$} \\
  \cline{2-6}
    $\nu$ & 1 & 2 & & 1 & 2     \\ \hline
   $\lam^*$ & $\displaystyle\frac{9 \sqrt{3}}{16}$ & 1 & & $\displaystyle\frac{9 \sqrt{3}}{16}$ & $\displaystyle\frac{\pi}{2}$ \\ \hline
\end{tabular}
\end{center}
\end{table}

For droplets with a non-trivial structure factor, the integral in eq.(\ref{onset}) can
be solved numerically. Whatsoever complicated were the structure factor $S(k)$, it is
anyway  obvious, that in this case $\lam^*$ is always greater, than the corresponding
value for point particles, since $S(k) < 1$ for non-zero $k$. In other words, the
droplets cannot be localized as easily as the point particles. If the density $n_i$ of
impurities is high, this is an odd conclusion, since it seems likely that a large
droplet gets stuck on a rugged potential landscape whereas a point particle may easier
elude the obstacles. On the other hand, given a small number of deep potential wells,
it is physically clear, that the point particle is trapped first, while a large
droplet is bound less effectively, in accord with the above conclusion. Hence, making
use of the terminology adopted in CDW theory \cite{fukuyama}, we can say, that the
present model corresponds the regime of strong pinning ($n_i$ low, $U_0$ large).

\begin{figure}[t]
\includegraphics[width=\linewidth,height=0.85\linewidth,angle=0]{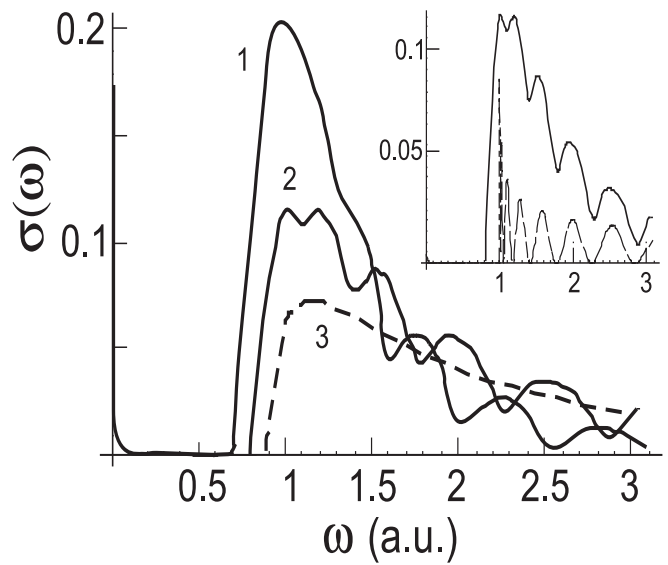}
\caption{The optical conductivity of logarithmic ($\nu = 2$) droplets with radius
$r_{out} = 15$ and different ring thickness parameter $\alpha$ for $\lam/\lam^* =
0.1$: 1 -- $\alpha = 0$, 2 -- $\alpha = 0.9$, 3 -- the spectrum of point Bose
particles \cite{epl}. Inset: the curve 2 together with the properly scaled structure
factor (dashed line) in rhs. of eq.(\ref{Mosc}).} \label{sw}
\end{figure}

To be specific, we consider the droplets, that have the form of the uniformly charged
rings with inner and outer radii denoted $r_{in}$, $r_{out}$, respectively. The charge
density is: $\rho(r) = (\theta(r - r_{in}) - \theta(r_{out} - r)) / S$, where $S = \pi
(r_{out}^2 - r_{in}^2)$ is the area of the ring, whence the form-factor (\ref{bessel})
is:
\begin{equation}\label{ring}
 \rho_k = \frac{2}{1 - \alpha^2}\left(\frac{J_1(k\,r_{out}) - \alpha J_1(\alpha\,k r_{out})}{k\,r_{out}} \right),
\end{equation}
where $\alpha = r_{in}/r_{out}$. The droplet is a disk for $\alpha = 0$, and a thin
ring for $\alpha$ close to unity. We show at Fig.1 (left panel) the dependence
$\lam^*(r_{out})$ for some values of $\alpha$ as derived from eq.(\ref{onset}) for
Coulomb droplets ($\nu = 1$) in the $\delta(r)$ impurity potential (other models,
listed in Table I, yield very similar results). As expected, the critical level
$\lam^*$ of the disorder, at which the glass transition takes place, increases rapidly
with the radius of the ring. It is also instructive to consider the dependence of
$\lam^*$ on the thickness $\alpha$ of the ring, its outer radius being fixed (Fig.1,
right panel). Noteworthy is the non-trivial variation of $\lam^*(\alpha)$ curves with
$r_{out}$. Two competing factors, the ring charge density and its spatial extension,
affect the coupling with impurity. As $\alpha$ increases from zero, $\lam^*$ begins to
grow since the charge moves on average to a greater distance from impurity and couples
weaker with it, albeit the charge density increases. However, this reasoning may
become wrong for large droplets, where the character of $\lam^*(\alpha)$ dependence
changes qualitatively. The behaviour of the curve 3 at right panel of Fig.1 means,
that the non-ergodicity transition for thin rings (large $\alpha$) of large enough
radius may occur at lower $\lam^*$, than for disks, i.e. these rings are more readily
trapped by impurities.

\begin{figure*}[t]
\begin{minipage}[b]{0.5\linewidth}
\includegraphics[width=\linewidth,angle=0]{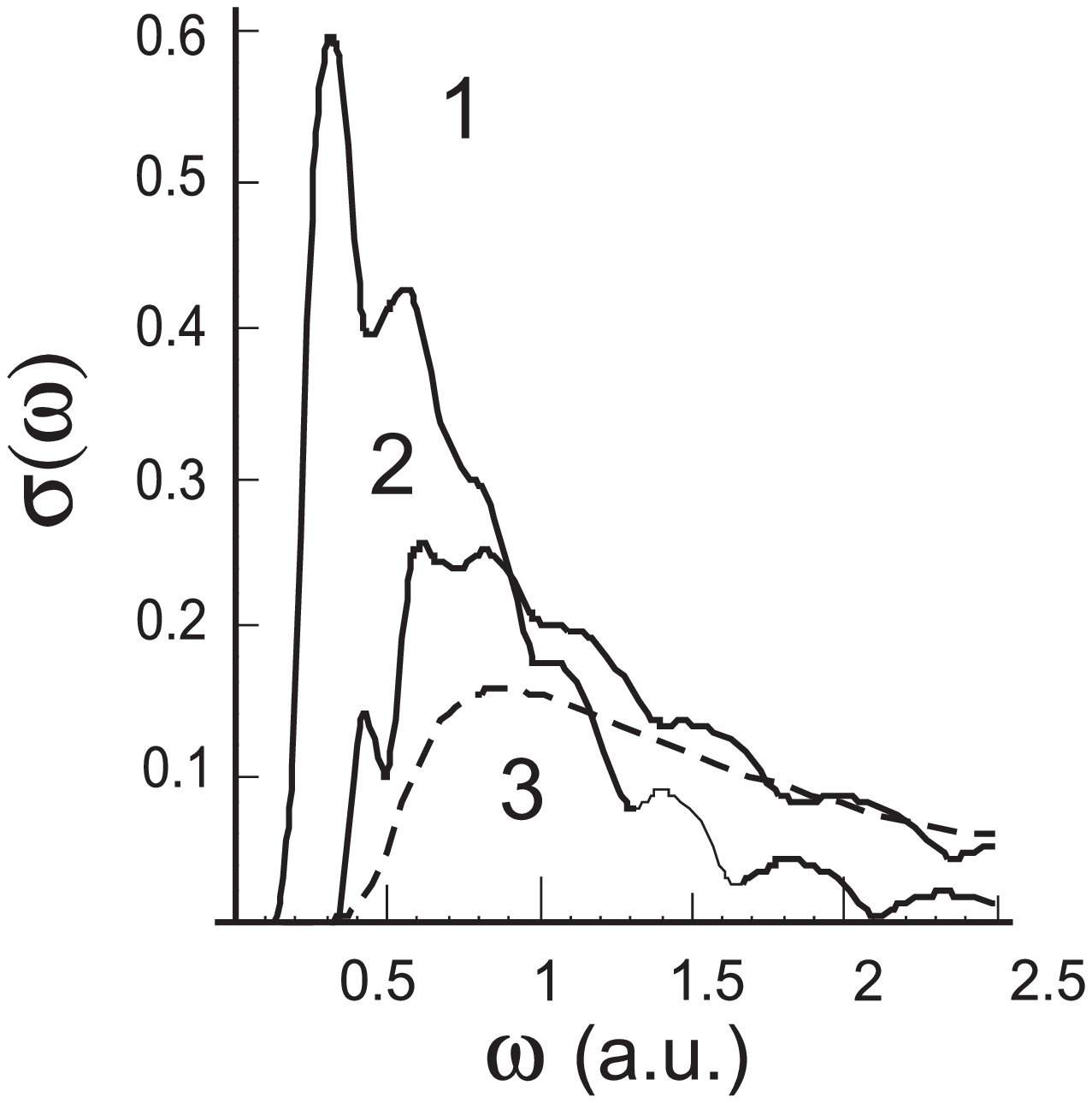}
\end{minipage}\hfill
\begin{minipage}[b]{0.5\linewidth}
\includegraphics[width=\linewidth,angle=0]{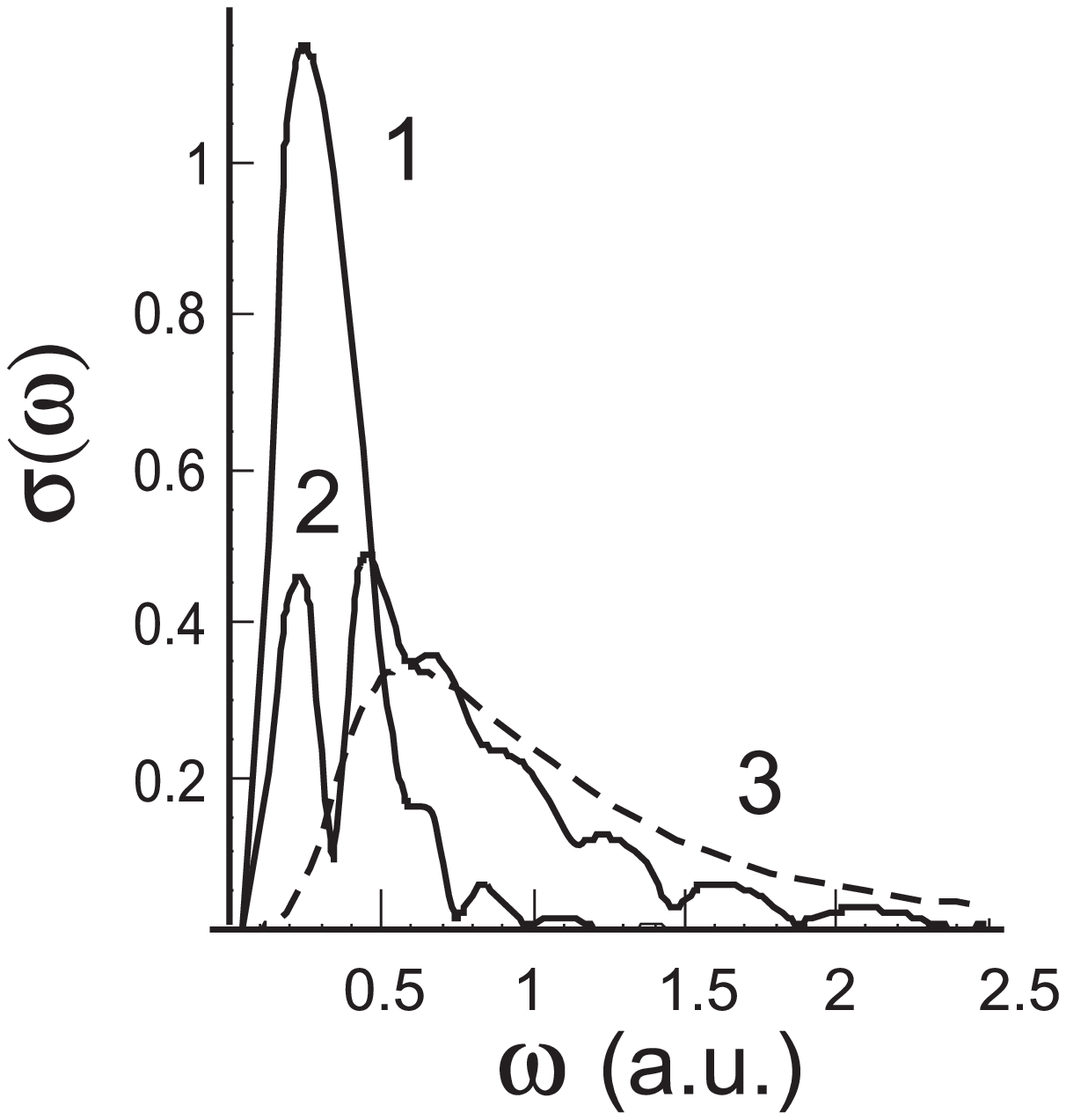}
\end{minipage}\hfill
\caption{Optical conductivity of Coulomb droplets with $r_{out} = 15$ and different
real-space structure in superfluid regime, $\lam/\lam^* = 0.25$. 1 -- $\alpha = 0$
(circles), 2 -- $\alpha = 0.9$ (rings), 3 -- point particles (no structure). Left
panel: short-range impurities ($\varphi = \delta(r)$).  Right panel: Coulomb
impurities ($\varphi = 1/r$).} \label{s23}
\end{figure*}

\subsection{Optical conductivity}

Now consider the effects of the structure factor (\ref{ring}) on the optical
conductivity spectra (\ref{drude}). Some typical examples of $\s(\w)$ spectrum are
shown at Figs.2,3. These results are obtained numerically from
eqs.(\ref{drude},\ref{m2},\ref{S},\ref{ring}) given $\lam/\lam^* < 1$ (superfluid
regime). The spectra display a broad asymmetric feature, related to the collective
(plasmon-like) density modes \cite{goldBose,epl}. As compared with the corresponding
spectra of point particles with the same $\lam/\lam^*$ ratio, that display a single
smooth feature (dashed curves at Figs.2,3), the spectra of the droplets are slightly
shifted to lower frequencies and exhibit a peculiar oscillating structure. The
oscillations become more pronounced with increasing radius of the droplets and (or)
inner-to-outer radius ratio $\alpha$.

The origin of the oscillations can be understood as follows. The main contribution to
the integral in eq.(\ref{m2}) comes from the vicinity of the wave numbers $k_i > 0$
($i = 1,2$):
\begin{equation}\label{poles}
k_i^4 + k_i^{2 - \nu} - \mbox{Re}\,\zeta = 0, \;\zeta = z (z + M(z)),
\end{equation}
where the integrand has the peaked structure. In particular, as far as $\lam \ll
\lam^*$, $|M(z)|$ is small and as a crude zero-order approximation, we may put $M = 0$
in rhs. of eq.(\ref{m2}), so that Im $\zeta$ is infinitesimal and the peaks are very
sharp. Then, the imaginary part of eq.(\ref{m2}) takes the form:
\begin{equation}\label{delta}
M^{\prime\prime} = \w\,\frac{8\,\lam}{\pi}\, \int \limits_0^{\infty} \sum_i R_i(k)\,S(k)\,\delta(k - k_i)\,dk,
\end{equation}
where $R_i(k)$ denotes the regular part of the integrand near $k = k_i$ and the
structure factor is not included in $R_i$. The spectral dependence of the memory
function and the optical conductivity (\ref{drude}) are thereby governed by the strongly
oscillating expression $S(k_i(\w))$. E.g., for $\nu = 2$ eqs.(\ref{poles},\ref{delta})
yield
\begin{equation}\label{Mosc}
M^{\prime\prime} = 2\,\lam\,\frac{\sqrt{\w^2-1}}{\w}\, S\left((\w^2 - 1)^{1/4}\right),\quad \w > 1.
\end{equation}
It may be seen at the inset of Fig.2, that the rhs. of eq.(\ref{Mosc}) reproduces
correctly the oscillations of the optical conductivity spectrum.

As the amount of the disorder $\lam/\lam^*$ increases, the oscillations of optical
conductivity become less pronounced. The point is that the imaginary part
$M^{\prime\prime}$ of the memory function and Im$\zeta$ increase with $\lam$ so that
the $\delta$-peaks in eq.(\ref{delta}) acquire a sizeable half-width $\Delta k$. Thus,
the contributions of the structure factor $S(k)$ add up from a range of wave numbers
$\left(k_i(\w) - \Delta\,k,\:k_i(\w) + \Delta\,k \right)$ and partially cancel each
other.

We have investigated also more complex axially symmetric charge distributions. All of
them reduce to a discrete or continuous series of concentric rings, considered above.
The resulting optical spectra are found to display a system of sharp and flat peaks of
the same origin as those depicted in Figs.2,3.

\section{Discussion}

The mode-coupling theory of the 2D weakly disordered Bose condensate have been first
proposed for the interpretation of the optical response of doped HTSC oxides in
Ref.\cite{goldHTS}. Despite a number of model assumptions, the results obtained
therein have been found to compare well with the properly scaled optical spectra of
YBa$_2$Cu$_3$O$_{7-\delta}$.

To our opinion, the approach considered in Ref.\cite{goldHTS} and in the present work
deserves attention in connexion with the far-IR features in optical conductivity
spectra of underdoped HTSC's, that are presumably related to the stripe excitations
\cite{Lucarelli} or Wigner crystal dynamics \cite{Hor}.
In particular, the in-plane optical conductivity of La$_{2-x}$Sr$_x$CuO$_4$ ($0 < x <
0.26$) is found to display strong resonant structure around $\sim 100$ cm$^{-1}$, that
consists of a main peak and a series of less pronounced peaks at higher frequencies
\cite{Lucarelli}. The authors interpret these features as the signature of the optical
response of charged stripes, extending over many ($\sim 10^2$) cells in the Cu-O
plane.

These results are only a part of a great body of findings \cite{mihailovic} obtained
with different techniques (neutron scattering \cite{Phsep}, microwave spectroscopy
\cite{Hor,Sridhar}, scanning tunnel microscopy \cite{pan}, etc.), that give convincing
evidence of the existence and an important role of a real-space inhomogeneity in the
physics of doped cuprates and related systems. The electronic inhomogeneity,
considered mainly in the form of density waves or stripes, suggests the percolative
nature of the transition between the insulating and superconducting regions of the
$T_c - x$ phase diagram of HTSC's \cite{PSeffect,Philips}. The latter implies the
existence of the nanoscale islands of superconducting phase above $T_c$, and it was
proposed, that ''the superconductivity arises predominantly in the charged stripes (or
domains)'' \cite{Markiewicz}.

The present model of Bose droplets may be relevant for the description of
electromagnetic response of such a system of local condensate patches. The droplets
emerge on Cu-O planes of layered HTSC's upon doping as the local bosons with charge Q
and effective mass $m^*$, delocalized within a region of a certain spatial extension,
that depends on the width of the potential wells created by doped impurities. The
droplet may be either rounded or ''stripe''-shaped -- in the latter case the droplets
should be randomly oriented so as to restore the isotropy of the system. At finite
frequencies the optical spectrum of the system consists of a broad asymmetric feature
(the plasmon wing) due to the collective plasmon-like excitations of the droplet
arrays. The latter is similar to the pinned CDW optical response, and it was recently
shown \cite{epl}, that the optical conductivity spectrum of the 2D bosons with
logarithmic interaction in random short-range potential ($\nu = 2$, $\varphi(r) =
\delta(r)$) exhibits just the same frequency dependence as the CDW in strong pinning
regime \cite{fukuyama}. However, as far as the Bose liquid remain confined within the
droplets, its real-space density is essentially nonuniform, and the effects of a
non-trivial structure factor are important. These manifest themselves in the strong
frequency modulation of the plasmon wing (Figs.2,3), governed by the structure factor
(eq.\ref{Mosc}).

We point out that the spectra at Fig.3 bear a good deal of resemblance with microwave
optical conductivity of lightly doped La$_{2-x}$Sr$_x$CuO$_4$ at $x \sim 0.1$, that is
supposedly dominated by charged stripes excitations \cite{Lucarelli}. The in-plane
concentration of the droplets is assumed to depend on the doping concentration $x$. Thus, $x = 0.1$ corresponds to $n_{2D} = 6.25\,10^{-12}$ cm$^{-2}$ (the area of the CuO$_4$ cluster being 16 \AA$^2$). Taking
$Q = 2 e$, $m^* = 3000\,m_e$ and the static dielectric constant $\eps = 40$ \cite{Hor}
($e$ and $m_e$ being the electron charge and mass, respectively), we obtain the unit
of energy (eq.\ref{units}) $\varepsilon_0 = 204.25$ cm$^{-1}$ and the unit of length
$\ell = 2 \pi \hbar/p_0 \simeq 15$ \AA. Hence, the oscillating structure of the
spectra at Fig.3 falls to the energy $\sim 100$ cm$^{-1}$ and the droplet size is
$\sim 225$ \AA, or about $40$ CuO$_4$ cells. These estimates are in agreement with the
energy of charged stripes excitations and the mean stripe size, reported in
\cite{Lucarelli}.

\begin{figure*}[t]
\begin{minipage}[b]{0.5\linewidth}
\includegraphics[width=\linewidth,angle=0]{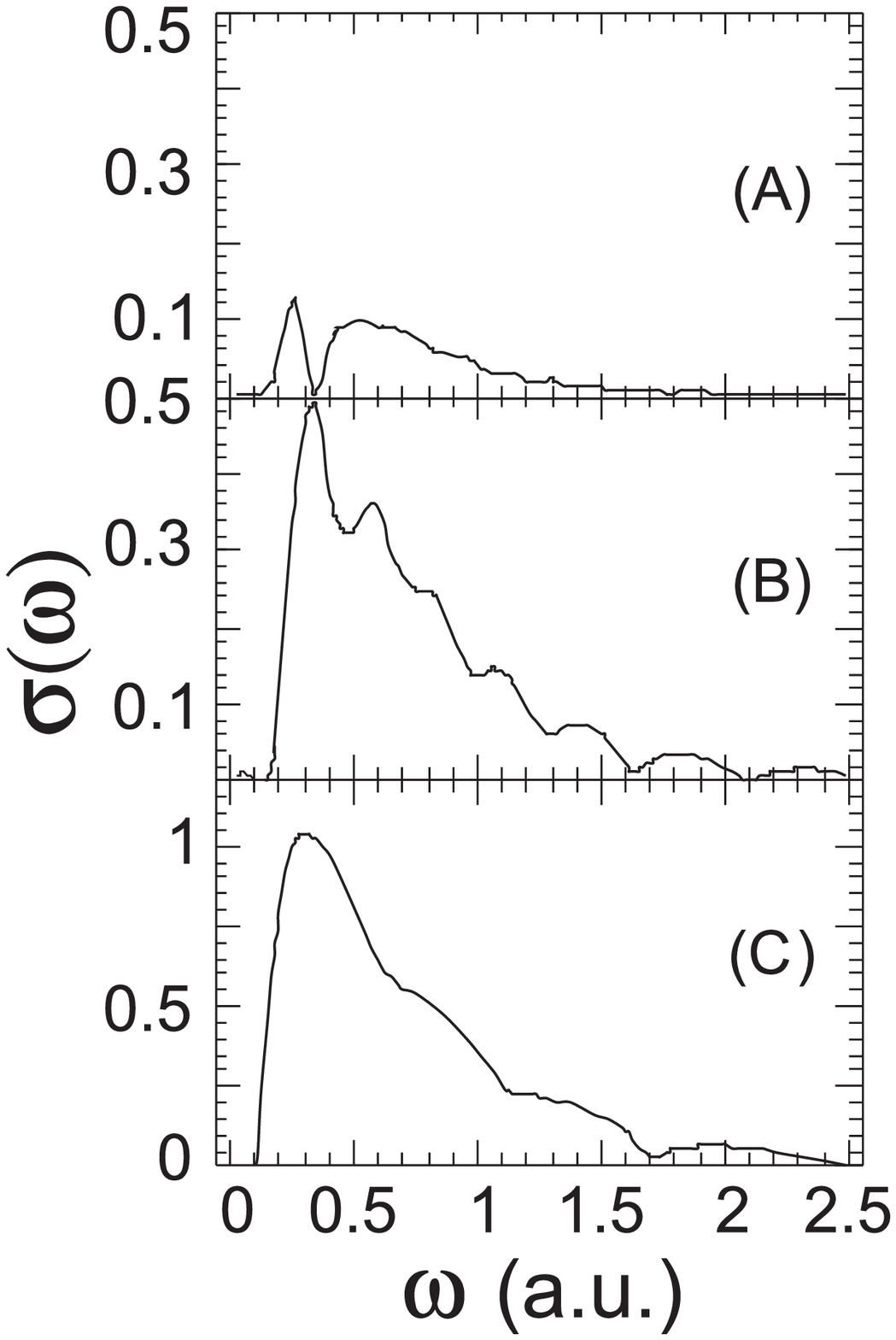}
\end{minipage}\hfill
\begin{minipage}[b]{0.5\linewidth}
\includegraphics[width=\linewidth,angle=0]{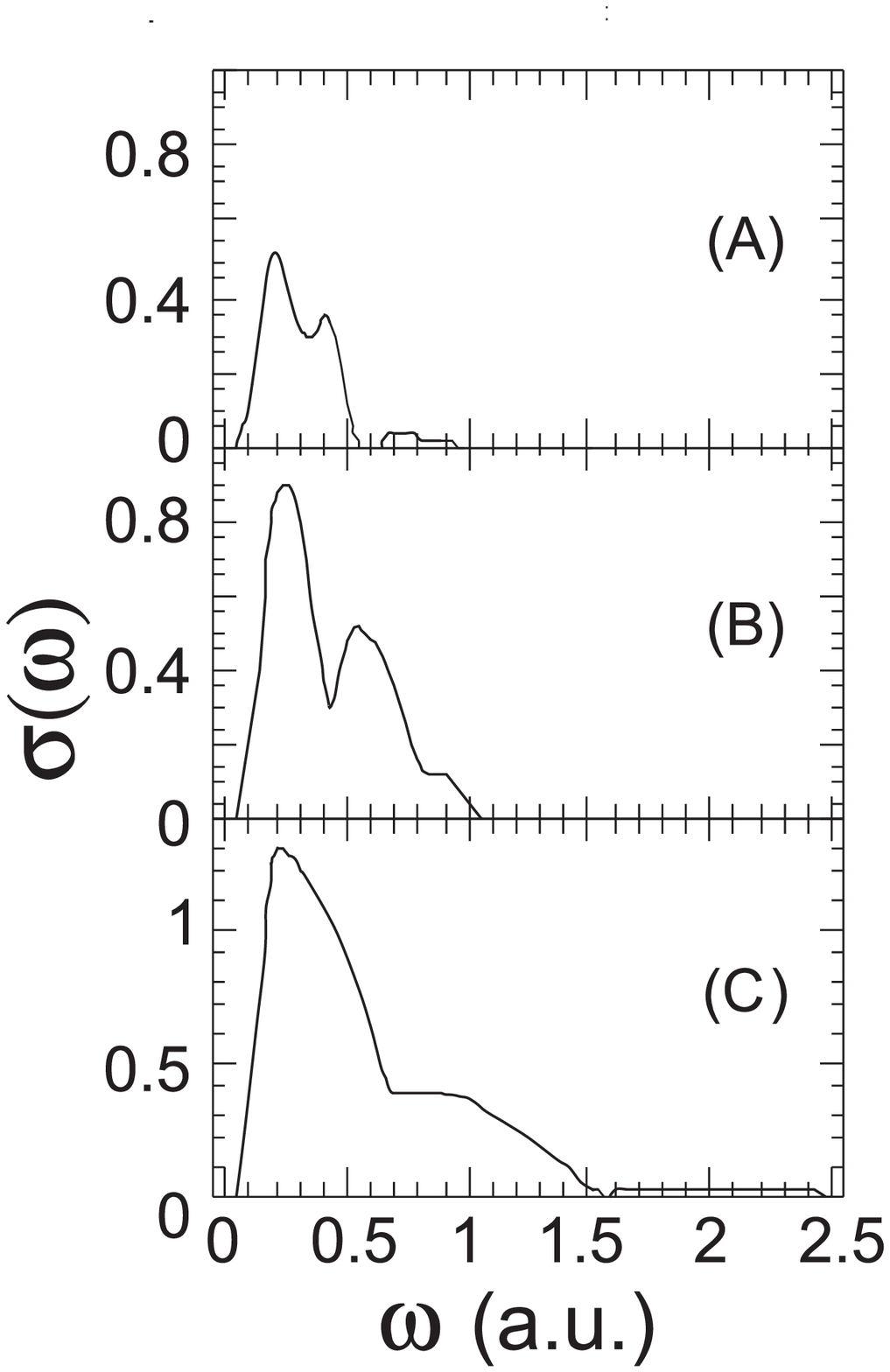}
\end{minipage}\hfill
\caption{Optical conductivity of Coulomb droplets at different parameters of the
disorder. Left: short-range impurities. A -- $\lam/\lam^* = 0.05$, B -- $\lam/\lam^* =
0.2$, C -- $\lam/\lam^* = 0.5$. Right: Coulomb impurities. A -- $\lam/\lam^* = 0.1$, B
-- $\lam/\lam^* = 0.25$, C -- $\lam/\lam^* = 0.5$.}
\end{figure*}

Assuming, that each impurity creates a droplet on CuO$_2$ plane and $n_{2D} \sim n_i$,
it follows from eq.(\ref{lamda}), that $\lam \sim 1/x$. However, it is not $\lam$
itself but the ratio $\lam/\lam^*$ that governs the shape of the optical conductivity
spectra, where $\lam^*$ (eq.\ref{onset}) depends on the structure of the droplet and
in particular, on its radius. A rigorous estimate of the droplet radius could be
obtained by minimization of free energy and is beyond the scope of the present work.
However, since we regard the droplet as a region of a coherent motion of the boson, it
is plausible, that its radius reduces with increasing doping concentration $x$, when
the system becomes inhomogeneous at smaller scale, whereby $\lam^*$ decreases with
$x$. As $\lam^*$ varies strongly with the droplet radius $r_{out}$ (see Fig.2), this
implies, that the ratio $\lam/\lam^*$ may grow with $x$, although $\lam$ and $\lam^*$
decrease.

At Fig.4 we show a series of spectra, that simulates the variation of the microwave
optical conductivity of La$_{2-x}$Sr$_x$CuO$_4$ \cite{Lucarelli} with doping. The left
panel corresponds to the disks ($\alpha = 0$) in short-range impurity potential,
$\varphi(k) = 1$. At early stage of doping (Fig.4A) there exists a small number of
large droplets with $r_{out} = 30.5$ and $\lam_A/\lam^*_A = 0.05$. In these conditions
the spectrum  consists of an almost flat background and a single distinct feature,
similarly to the optical conductivity of LSCO at $x \sim 0.07$. At Fig.4B it evolves
into the wing with a well-pronounced structure, as discussed in previous section. This
spectrum corresponds to $x \sim 0.15$ and is obtained with $\lam^B : \lam^A \simeq
0.07 : 0.15$ and reduced droplet radius $r_{out} = 15$, that provides $\lam_B/\lam^*_B
= 0.2$. The spectrum at Fig.4C corresponds to $x \sim 0.26$ and is obtained given
$\lam_C : \lam_B \simeq 0.15 : 0.26$ and still smaller droplets with $r_{out} = 8$, so
that $\lam_C/\lam^*_C = 0.5$. It is seen, that the spectrum broadens, loses its fine
structure and its magnitude grows sizeably, in agreement with experiment. Overall, a
qualitative description of all essential points in doping dependence of the optical
conductivity of LSCO, reported in Ref. \cite{Lucarelli}, becomes possible. Similarly,
the curves at right panel of Fig.4 have been obtained for rings with $\alpha = 0.5$
given the following parameters: A -- $\lam/\lam^* = 0.1$ , $r_{out} = 36$, B --
$\lam/\lam^* = 0.5$, $r_{out} = 15$, C -- $\lam/\lam^* = 0.5$, $r_{out} = 7$. Again,
the values of $\lam$ in these examples meet the proportions: $\lam^A : \lam^B \simeq
0.15 : 0.07$ and $\lam^C : \lam^B \simeq 0.26 : 0.15$.

The experimental findings of Ref.\cite{Lucarelli} have been examined also within the
diagrammatic theory of charged stripes \cite{benfatto}. In agreement
with the CDW theory by Fukuyama and Lee \cite{fukuyama}, the authors have found the optical
conductivity of an isolated stripe to display a peak at a pinning frequency $\w_0$,
whereas the the spectra of an array of interacting stripes show a second peak at a
frequency $\w_U$, that depends on the interaction strength. Physically, the peak at
$\w_0$ corresponds to the plasmon wing (that is not necessarily broad and may look
like a single resonance, as at right panel of Fig.3), while the peak at $\w_U$ has no
direct counterpart in our model. However, the observed microwave spectra in cuprates
generally show a much more complex structure with a number of resonances.

In conclusion, we have studied the optical response of the liquid of interacting 2D
Bose droplets in presence of a random impurity potential -- either Coulomb or
short-range. Each droplet is a charged Bose particle, delocalized with a charge
density $\rho(r)$ in a certain spatial region (disk or ring in the present work). As
discussed in Sec.IIIA, the behaviour of this system is similar to that of CDW in
strong pinning regime (a dilute distribution of strong pinning centers). The optical
conductivity spectrum of the model consists of a broad asymmetric feature (plasmon
wing) with a rather complex oscillating shape. The oscillations are intimately related
to the behaviour of the droplet structure factor $S = \rho(k)^*\,\rho(k)$. The main
qualitative conclusion to be drawn from our analysis is that the optical response is
not only related to the peculiarities of the energy level distribution, but may also
reflect the details of the real-space inhomogeneity of the system, such as the charge
density within the droplets of a (super)conducting phase. The model can possibly
describe the microwave optical response of doped cuprates, where the signature of the
collective mode excitations (density waves, stripes) are observed.

\begin{acknowledgments}
Thanks are due to Michele Cappellari for making available his package for
Mathematica computational system. The work is partially supported by Grant 04-02-96068
RFBR URAL 2004.
\end{acknowledgments}


\begin{thebibliography}{99}
\bibitem{note}%
 The details for $\nu = 2$ are given in Ref.\cite{epl} and the derivation for $\nu = 1$ is similar.
\end{thebibliography}
\end{document}